\documentclass[twoside,12pt]{article}
\UseRawInputEncoding
\usepackage{indentfirst}
\usepackage{epsfig}
\usepackage{latexsym, graphicx}
\usepackage{amsmath, amssymb, bm}
\usepackage{hyperref}
\usepackage{longtable}
\usepackage{color}
\usepackage{float}

\newcommand{\mi}{\mathrm{i}}
\newcommand{\me}{\mathrm{e}}
\newcommand{\md}{\mathrm{d}}
\newcommand{\mT}{\mathrm{T}}

\begin{document}

%-------------------  First Head  -----------------------------------------
\thispagestyle{empty} \vspace*{0.8cm}\hbox
to\textwidth{\vbox{\hfill\huge\sf Commun. Theor. Phys.\hfill}}
\par\noindent\rule[3mm]{\textwidth}{0.2pt}\hspace*{-\textwidth}\noindent
\rule[2.5mm]{\textwidth}{0.2pt}

%=================== Text begin here =============================================

\begin{center}
\LARGE\bf Superconducting gap ratio from strange metal phase in the absence of quasiparticles
\end{center}

\footnotetext{\hspace*{-.45cm}\footnotesize $^*$Corresponding author, E-mail: gexh@shu.edu.cn}

\begin{center}
\rm Wenhe Cai$^{\rm 1}$\ and  \ Xian-Hui Ge$^{\rm 1,2,*}$
\end{center}

\begin{center}
\begin{footnotesize} \sl
${}^{\rm 1}$ Department of Physics, Shanghai University, Shanghai 200444, China\\
${}^{\rm 2}$ Shanghai Key Laboratory of High Temperature Superconductors, Department of Physics, Shanghai University, Shanghai 200444, China\\
\end{footnotesize}
\end{center}

\vspace*{2mm}

\begin{center}
\begin{minipage}{15.5cm}
\parindent 20pt\footnotesize
A lattice model for strongly interacting electrons motivated by a rank-3 tensor model provides a tool for understanding the pairing mechanism of high-temperature superconductivity. This SYK-like model describes the strange metal phase in the cuprate high temperature superconductors. Our calculation indicates that the superconducting gap ratio in this model is higher than the ratio in the BCS theory due to the coupling term and the spin operator. Under certain conditions, the ratio also agrees with the BCS theory. Our results relate to the case of strong coupling, so it may pave the way to gaining insight into the cuprate high temperature superconductors.
\end{minipage}
\end{center}

\begin{center}
\begin{minipage}{15.5cm}
\begin{minipage}[t]{2.3cm}{\bf Keywords:}\end{minipage}
\begin{minipage}[t]{13.1cm}
holography and condensed matter physics, superconducting gap ratio, strange metal
\end{minipage}\par\vglue8pt

\end{minipage}
\end{center}
\section{Introduction}

The strange metal phase do not have long-lived quasiparticles. A toy model called Sachdev-Ye-Kitaev (SYK) model captures the feature of the strange metal phase, such as the absence of quasiparticles. The SYK model is a disordered and strongly-coupled quantum system composed by $N$ Majorana fermions with Gaussian-distributed random coupling \cite{SY,K,Maldacena:2016upp}. The connection between the SYK model and the gravity theory with a near-horizon $\rm AdS_2$ geometry could be obtained in \cite{Maldacena:2016hyu,Polchinski:2016xgd,Jensen:2016pah}. Various applications of SYK model have been presented, such as topological SYK model \cite{ZZ}, SYK-like models \cite{Krishnan:2016bvg,DJY,Bonzom:2017pqs,Krishnan:2017ztz,Peng:2018zap,Krishnan:2018jsp,Gross:2016kjj,Chaturvedi:2018uov}, transport \cite{Jian:2017unn,Davison:2016ngz,Cai:2017vyk,Ge:2018lzo}, SYK spectral density \cite{Jia:2018ccl,a:2018kvh,Garcia-Garcia:2016mno,Garcia-Garcia:2017pzl,Das:2017hrt,Garcia-Garcia:2018pwt}, supersymmetric SYK model \cite{Fu:2016vas,Peng:2016mxj,Li:2017hdt,Hunter-Jones:2017raw,Narayan:2017hvh}, complexity\cite{Sun:2019yps},quantum choas \cite{Jensen:2016pah,MSS}, the higher dimensional generalization \cite{Khveshchenko:2017mvj,Murugan:2017eto,Narayan:2017qtw,Berkooz:2016cvq} and the bulk gravity dual of SYK models \cite{Qi:2018rqm,Das:2017wae,Li:2018omr,Jian:2017tzg,Cai:2017nwk}.

Recently, it was found that the SYK model could be a powerful method to study strong coupling superconductivity. Actually, the conventional superconductors can be described by the BCS theory. However, the BCS theory is not capable of explaining the high temperature superconductor. The theories of high-temperature superconductivity extend the canonical BCS theory to strong electron-phonon coupling. There are some progresses on high-temperature superconductivity within the framework of SYK dots \cite{Patel:2017mjv,Patel:2018rjp,Garcia-Garcia:2018ruf,Chew:2017xuo}. The single particle phase has been investigated \cite{Patel:2018zpy}. There is a finite-temperature crossover to an incoherent metal (IM) and the marginal-Fermi liquid (MFL) \cite{Patel:2017mjv} or crossover to MFL and non-Fermi liquid (NFL) \cite{Chowdhury:2018sho} in some lattice models. The SYK model realizes a gapless non-fermi liquid, and it violates the ratio between the zero temperature gap and the critical temperature which predicted by BCS mean-field theory \cite{Patel:2018rjp}.

The  q-body ($q$ is even) SYK hamiltonian is $H = (\mi)^{ q\over 2 } \sum_{1\leq i_1 < i_2 < \cdots < i_q \leq N }    j_{i_1 i_2\cdots i_q} \psi_{i_1} \psi_{i_2} \cdots \psi_{i_q}$, where $J_{i_1,...,i_q}$ are Gaussian random variables. In the case $q>2$, this model describes a non-Fermi liquid without quasiparticles \cite{K,Maldacena:2016hyu}. Although the SYK model describes a non-Fermi liquid state, it actually has marginally relevant paring instability just like the ordinary Fermi liquid state in some previous works \cite{Bi:2017yvx,Luo:2017bno}. In the case $q=2$, the random mass-like Hamiltonian can be diagonalized. The case of two-body interactions is trivial since free fermion terms dominate at low energies \cite{Eberlein:2017wah}.

As a candidate theory in \cite{WCJYX}, the authors propose a lattice model for strongly interacting electrons motivated by the recently developed ``tetrahedron" rank-3 tensor model that mimics much of the physics of the SYK model (See more details in \cite{Witten:2016iux,Klebanov:2016xxf,Gurau:2009tw}). This model can explain some of the strange metal phase in the cuprate high temperature superconductors. The single particle Green's function of this lattice model in the large $N$ limit is identical to the disorder-averaged Green's function of the SYK model. The lattice model leads to a fermion pairing instability just like the BCS instability. The system could form SP(M) spin singlet fermion pairings. Within the framework of their model, we further study the superconducting gap ratio in the absence of quasiparticles. Our scenario is analogous to Cooper's argument. We explore a pairing mechanism in this $(2+1)$-dimensional lattice model for strongly interacting electrons.

The paper is organized as follows, in section 2, we construct symplectic group singlet pairs between fermions in the transverse momentum space and the corresponding microscopic model. Then, we derive equations for the correlation functions. In section 3, we investigate the gap function, the transition temperature and the ratio. We also evaluate the influence of the attractive term and spin term, and compare our results with the BCS theory. The section 4 is the summary and discussion.

\section{SP(M) singlet pairs and the microscopic model}

In this section, we construct singlet pairs between only two sites and briefly review the microscopic lattice model. We first introduce a 2M-component fermion basis on site 1 and site 2,
\begin{equation}
     \Psi=\big(c_{1,\alpha},\ c^\dag_{2,\alpha}\big)^{\mT}\,.
\end{equation}
The $2M\times 2M$ Green's function matrix is given by
\begin{equation}
     -\langle T_\tau\Psi(\tau)\Psi^\dag(0)\rangle=\left(\begin{array}{cc}
                                  -\langle T_\tau c_{1,\alpha}(\tau)c^\dag_{1,\beta}(0)\rangle &  -\langle T_\tau c_{1,\alpha}(\tau)c^\dag_{2,\beta}(0)\rangle\\
                                   -\langle T_\tau c_{2,\alpha}(\tau)c^\dag_{1,\beta}(0)\rangle &  -\langle T_\tau c_{2,\alpha}(\tau)c^\dag_{2,\beta}(0)\rangle\\
                               \end{array}
                            \right)\,.\nonumber
\end{equation}
Then we consider a general dimer of site ($\bm{i},\bm{j}$). Here $\Delta_{\bm{i},\bm{j}}=J_{\alpha\beta}c_{\bm{i},\alpha}c_{\bm{j},\beta}$ is an SP(M) spin singlet fermion pairings on nearest neighbor links $\langle\bm{i},\bm{j}\rangle$.
Motivated by the observation that the symplectic group SP(M) allows fermions to form singlet pairs \cite{FDC,RS}, we define
\begin{eqnarray}
     \mathcal{G}_{\bm{i},\bm{i}}(\tau)&=&-\langle T_\tau \delta_{\alpha\beta}c_{\bm{i},\alpha}(\tau)c^\dag_{\bm{i},\beta}(0)\rangle\,,\\
     \mathcal{F}_{\bm{i},\bm{j}}(\tau)&=&\langle T_\tau J_{\alpha\beta}c_{\bm{i},\alpha}(\tau)c_{\bm{j},\beta}(0)\rangle\,,\\
     \mathcal{F}^\dag_{\bm{i},\bm{j}}(\tau)&=&\langle T_\tau J_{\alpha\beta}c^\dag_{\bm{i},\alpha}(\tau)c^\dag_{\bm{j},\beta}(0)\rangle\,.
\end{eqnarray}
It is similar to the cooper pair in the neighbor site $\langle T_\tau c_{-\mathbf{p},\uparrow}(\tau)c_{\mathbf{p},\downarrow}(0)\rangle$.
The creation operator in Fourier space is $c^\dag_{\bm{j},\alpha}=\sum_{\bm{p}}\me^{\mi\bm{j}\cdot\bm{p}}c^\dag_{\alpha,\bm{p}}$. Here site indices $\bm{i}=(i_x,i_y)$ and the conjugate momentum $\bm{p}=(p_x,p_y)$ are two-dimensional vectors. Thus, the Fourier transformations of the pair are
\begin{eqnarray}
     c_{\bm{j},\alpha}(\tau)c^\dag_{\bm{j},\beta}(0)&=&\sum_{\bm{p},\mathbf{p'}}\me^{-\mi\bm{j}\cdot(\bm{p}-\bm{p'})}c_{\bm{p},\alpha}(\tau)c^\dag_{\bm{p'},\beta}(0)\,,\\
     c_{\bm{i},\alpha}(\tau)c_{\bm{j},\beta}(0)&=&\sum_{\bm{p},\bm{p'}}\me^{-\mi(\bm{i}\cdot\bm{p}+\bm{j}\cdot\bm{p'})}c_{\bm{p},\alpha}(\tau)c_{\bm{p'},\beta}(0)\,,\\
     c^\dag_{\bm{i},\alpha}(\tau)c^\dag_{\bm{j},\beta}(0)&=&\sum_{\bm{p},\bm{p'}}\me^{\mi(\bm{i}\cdot\bm{p}+\bm{j}\cdot \bm{p'})}c^\dag_{\bm{p},\alpha}(\tau)c^\dag_{\bm{p'},\beta}(0)\,.
\end{eqnarray}
By introducing the momentum and the hopping term, we modify the interacting electron Hamiltonian in \cite{WCJYX} as follows,
\begin{align}
  H=&\sum_{\bm{q}\bm{p}\bm{p'}}\tilde{U}(\bm{p})c^\dag_{\sigma,\bm{q}}c_{\sigma,\bm{q}+\bm{p}'}c^\dag_{\gamma,\bm{p}}c_{\gamma,\bm{p}-\bm{p}'}+\sum_{\bm{p}}\xi_{p}c^\dag_{\bm{p},\sigma}c_{\bm{p},\sigma}\nonumber\\
    &-\frac{1}{4}J\sum_{\bm{p},\bm{q}}c^\dag_{\bm{p},\sigma}c_{\bm{p},\sigma}c^\dag_{\bm{q},\gamma}c_{\bm{q},\gamma}+\frac{1}{2}J\sum_{\bm{p},\bm{q}}c^\dag_{\bm{p},\alpha}\sigma_{\alpha\beta}c_{\bm{p}\bm{p},\beta}c^\dag_{\bm{q},\beta}\sigma_{\beta\alpha}c_{\bm{q},\alpha}\nonumber\\
    &+K\sum_{\bm{q}\bm{p}\bm{p}'}\bigg(\epsilon_{\alpha\beta}\epsilon_{\gamma\sigma}c^\dag_{\alpha,\bm{p}+\bm{q}}c^\dag_{\beta,\bm{p}'-\bm{q}}c_{\gamma,\bm{p}'}c_{\sigma,\bm{p}}+\rm{H.C.}\bigg)\,.
\end{align}
We have set the volume $\upsilon=1$ for simplification. $\hat{n}_{\bm{i}}=\hat{n}_{\bm{i},\uparrow}+\hat{n}_{\bm{i},\downarrow}$ is the total electron number on site $\bm{i}$. $\vec{S}_{\bm{i}}=\frac{1}{2}c^\dag_{\bm{i}}\vec{\sigma}c_{\bm{i}}=\frac{1}{2}c^\dag_{\bm{i},\alpha}\sigma_{\alpha\beta}c_{\bm{i},\beta}$ is the spin operator, and $\vec{S}_{\bm{i}}\cdot\vec{S}_{\bm{j}}=\frac{1}{2}\vec{S}_{\alpha\beta,\bm{i}}\vec{S}_{\beta\alpha,\bm{j}}$. $\xi_q$ is the energy of the single particle which hoppings between the two sublattices as perturbations. $K$ satisfies
\begin{center}
    $K\left\{
              \begin{array}{ll}
                    <0, & |\xi_q|<\omega_D\,,\\
                    =0, & |\xi_q|>\omega_D\,.
              \end{array}
        \right.$
\end{center}
Here $\omega_D$ is the Debye energy. The term with the coupling $K$ takes a spin singlet pair of electrons on two diagonal sites $\bm{j},\bm{j}+\hat{x}+\hat{y}$ of a
plaquette to the two opposite diagonal sites $\bm{j}+\hat{x},\bm{j}+\hat{y}$ of the same plaquette. The perturbation with coefficient $K$ forms SP(M) spin singlet fermion pairings. Only when $\tilde U=K=\pm J/2$, the interacting electron model in \cite{WCJYX} is equivalent to a tetrahedron model with three indices: the SP(M) spin, the $x$ coordinate, and $y$ coordinate.
\begin{align}
  &\frac{g}{N_aN_bN_c}J_{c1c1'}J_{c2c2'}c^\dag_{a1b1c1}c^\dag_{a2b2c1'}c_{a1b2c2}c_{a2b1c2'}\nonumber\\
  \sim&\frac{g\eta_{r,r'}}{N\sqrt{M}}J_{\alpha\beta}J_{\gamma\sigma}c^\dag_{jx,jy,\alpha}c^\dag_{jx+r,jy+r',\beta}c_{jx,jy+r',\gamma}c_{jx+r,jy,\sigma}\,,\nonumber
\end{align}
where $g$ is  the same order as the coupling $J$. The total symmetry of this model is $U(N_a)\times U(N_b)\times SP(N_c)$.

\section{The gap function and the transition temperature}

\begin{figure}[H]
\centerline{\includegraphics[width=7.5cm]{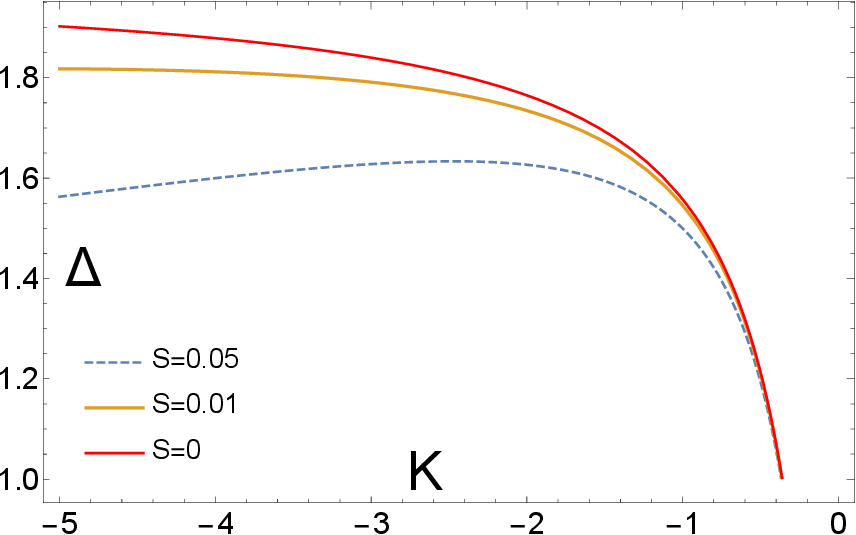}\quad\includegraphics[width=7.5cm]{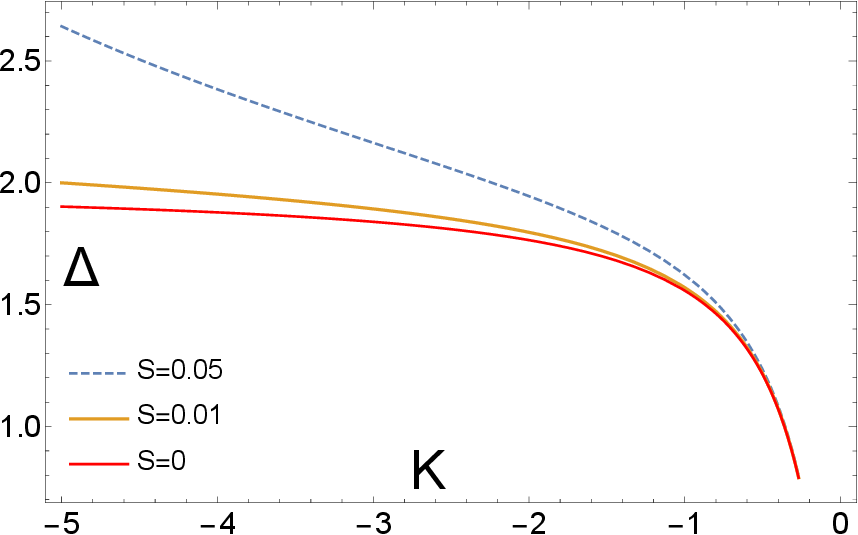}}
\caption{The relation between the gap $\Delta$ and the coupling $-5<K<1/2$ with different $S=\langle T_\tau\sigma c^\dag Sc^\dag\rangle= 0, 0.01, 0.05$ represented by red, purple, dashed respectively. The figure on the left corresponds to the case of $U=K=J/2$. The figure on the right corresponds to the case of $U=K=-J/2$. The gap changes abruptly when $K$ goes from negative to zero.}
\label{fg:Figure 1}
\end{figure}

\begin{figure}[H]
\centerline{\includegraphics[width=10cm]{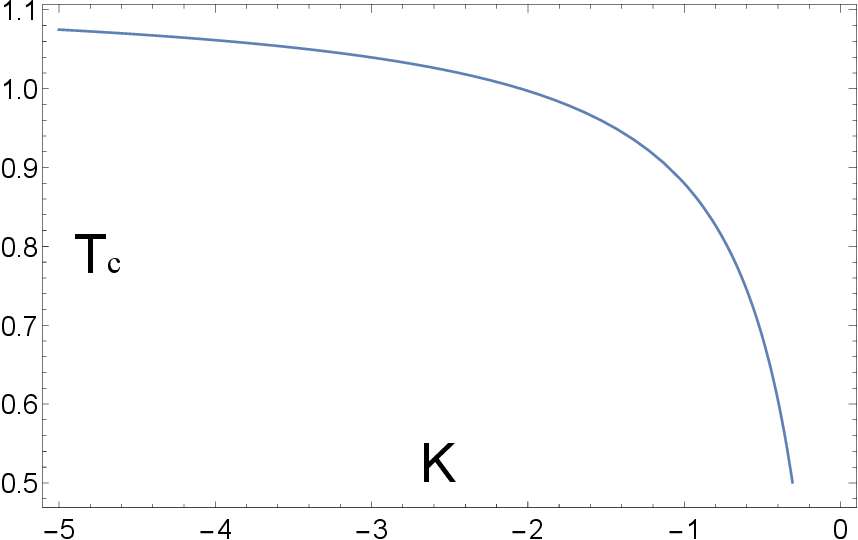}}
\caption{The figure shows the relation between the transition temperature $T_c$ and $-5<K<1/2$ in the case of $U=K=\pm J/2$. The transition temperature changes abruptly when $K$ goes from negative to positive.}
\label{fg:Figure 2}
\end{figure}

\begin{figure}[H]
\centerline{\includegraphics[width=7.5cm]{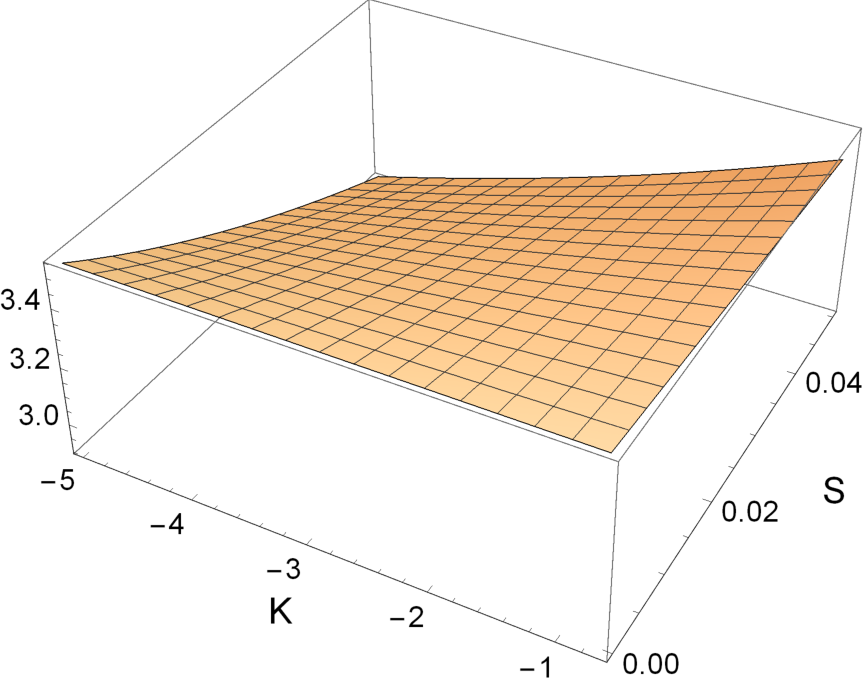}\quad\includegraphics[width=7.5cm]{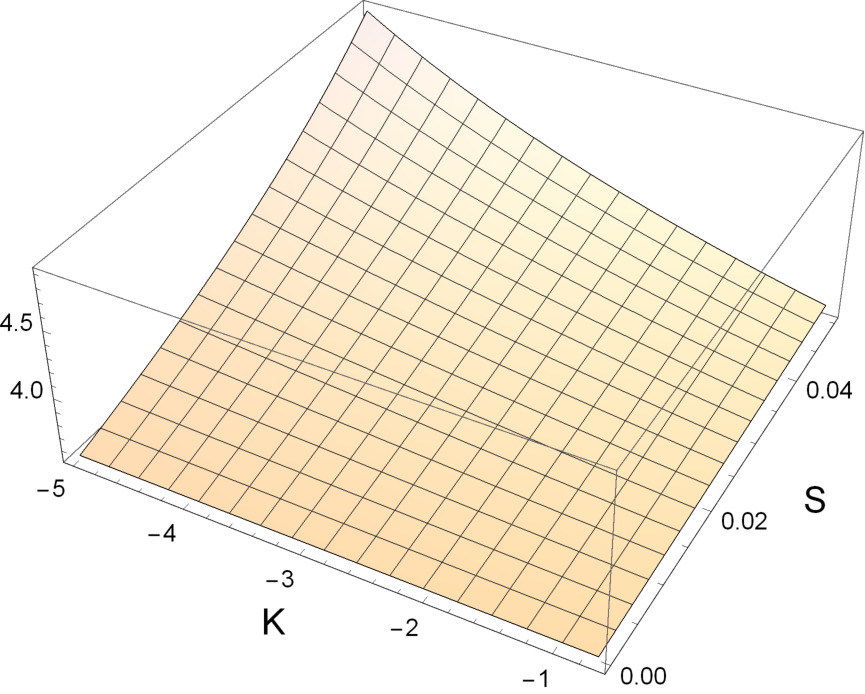}}
\caption{The dependence of $\frac{2\Delta}{T_c}$ on $K$ and S. The figure on the left shows that the ratio decrease as $K$ decrease and S increase in the case of $U=K=J/2$. The figure on the right shows that the ratio increase as $K$ decrease and S increase in the case of $U=K=-J/2$.}
\label{fig:Figure 3}
\end{figure}

As we are going to evaluate the gap ratio, let us first consider the time development
\begin{align}
  \frac{\md}{\md\tau}c_{\alpha,\bm{p}}(\tau)&=\big[H,c_{\alpha,\bm{p}}\big]\nonumber\\
                                         &=-2\sum_{\bm{q},\bm{p}'}\tilde{U}c^\dag_{\sigma,\bm{q}}c_{\sigma,\bm{q}+\bm{p}'}c_{\alpha,\bm{p}-\bm{p}'}+\frac{J}{2}c_{\alpha,\bm{p}}c^\dag_{\gamma,\bm{q}}c_{\gamma,\bm{q}}\nonumber\\
                                         &\ -\xi_{p}c_{\alpha,\bm{p}}-J\sigma_{\alpha\beta}c_{\beta,\bm{p}}\tilde{\vec{S}}+ 4K\sum_{\bm{q},\bm{p}'}\epsilon_{\alpha\beta}\epsilon_{\gamma\sigma}c^\dag_{\beta,\bm{p}'\bm{q}}c_{\gamma,\bm{p}'}c_{\sigma,\bm{p}-\bm{q}}\,,\label{eq:Dc1}\\
  \frac{\md}{\md\tau}c^\dag_{\alpha,\bm{p}}(\tau)&=\big[H,c^\dag_{\alpha,\bm{p}}\big]\nonumber\\
                                              &=2\sum_{\bm{q},\bm{p}'}\tilde{U}c^\dag_{\sigma,\bm{q}}c_{\sigma,\bm{q}+\bm{p}'}c_{\alpha,\bm{p}-\bm{p}'}-\frac{J}{2}c_{\alpha,\bm{p}}c^\dag_{\gamma,\bm{q}}c_{\gamma,\bm{q}}\nonumber\\
                                              &\ +\xi_{p}c^\dag_{\alpha,\bm{p}}+J\sigma_{\beta\alpha}c^\dag_{\beta,\bm{p}}\tilde{\vec{S}}+ 4K\sum_{\bm{q},\bm{p}'}\epsilon_{\gamma\beta}\epsilon_{\alpha\sigma}c_{\gamma,\bm{p}'+\bm{q}}c^\dag_{\beta,\bm{p}'-\bm{q}}c_{\sigma,\bm{p}'}\,,\label{eq:Dc2}
\end{align}
where $\tilde{\vec{S}}=\frac{1}{2}\sum_{\bm{q}}c^\dag_{\bm{q},\alpha}\sigma_{\alpha\beta}c_{\bm{q},\beta}$ and $\bm{q}=(0,p'_y-p_y)$. The equations for the correlation functions
\begin{eqnarray}
\mathcal{G}(\bm{p},\tau)&=&-\langle T_\tau\delta_{\alpha\beta}c_{\bm{p},\alpha}(\tau)c^\dag_{\bm{p},\beta}(0)\rangle\,,\nonumber\\
\mathcal{F}^\dag(\bm{p},\tau)&=&\langle T_\tau J_{\alpha\beta}c^\dag_{\bm{p},\alpha}(\tau)c^\dag_{-\bm{p},\beta}(0)\rangle\,,\nonumber
\end{eqnarray}
are determined by
\begin{eqnarray}
     \frac{\partial}{\partial\tau}\mathcal{G}(\bm{p},\tau)&=&-\delta(\tau)-\langle T_\tau\delta_{ab}\big[ \frac{\partial}{\partial\tau}c_{\bm{p},a}(\tau)\big]c^\dag_{\bm{p},b}(0)\rangle\,,\\
     \frac{\partial}{\partial\tau}\mathcal{F}^\dag(\bm{p},\tau)&=&\langle T_\tau J_{ab}\big[\frac{\partial}{\partial\tau}c^\dag_{\bm{p},a}(\tau)\big]c^\dag_{-\bm{p},b}(0)\rangle\,.
\end{eqnarray}
Combined with the results (\ref{eq:Dc1})(\ref{eq:Dc2}) and the gap function
\begin{equation}
  \Delta(\bm{p})=-4\sum_{\bm{q}}K\mathcal{F}^\dag(\bm{p}-\bm{q},\tau=0)\,,
\end{equation}
 the derivative of the equation for the correlation function after Fourier transforming is given as,
\begin{align}
  &(\mi p_n-\xi_p)\mathcal{G}(\bm{p},\mi p_n)+\Delta(\bm{p})\mathcal{F}^\dag(\bm{p},\mi p_n)+J\langle T_\tau \sigma_{a\beta}c_{\bm{p},\beta}\tilde{\vec{S}}c^\dag_{\bm{p},a}\rangle\nonumber\\
  &+2\sum_{\bm{q},\bm{p}'}\tilde{U}c^\dag_{\sigma,\bm{q}}c_{\sigma,\bm{q}+\bm{p}'}c_{a,\bm{p}-\bm{p}'}c^\dag_{a,\bm{p}}-\frac{J}{2}\sum_{\bm{q}}c_{a,\bm{p}}c^\dag_{\gamma,\bm{q}}c_{\gamma,\bm{q}}c^\dag_{a,\bm{p}}=1\,,\label{eq:G}\\
  &(\mi p_n+\xi_p)\mathcal{F}^{\dag}(\bm{p},\mi p_n)+J_{ab}\epsilon_{ab}\Delta^\dag(\bm{p})\mathcal{G}(\bm{p},\mi p_n)+J\langle T_\tau J_{ab}\sigma_{\beta a}c^\dag_{\bm{p},\beta}\tilde{\vec{S}}c^\dag_{-\bm{p},b}\rangle\nonumber\\
  &+2\sum_{\bm{q},\bm{p}'}J_{ab}\tilde{U}c^\dag_{\sigma,\bm{q}}c_{\sigma,\bm{q}-\bm{p}'}c_{a,\bm{p}+\bm{p}'}c^\dag_{b,-\bm{p}}-\frac{J}{2}\sum_{\vec{q}}J_{ab}c_{a,\bm{p}}c^\dag_{\gamma,\vec{q}}c_{\gamma,\bm{q}}c^\dag_{b,-\bm{p}}=0\,.\label{eq:F}
\end{align}
After the combination of the two equations, we simplify the final results as follows,
\begin{align}
  \mathcal{G}(\bm{p},\mi p_n)&=-\bigg[p^2_n+\xi^2_p+\Delta(\bm{p})J_{ab}\epsilon_{ab}\Delta^\dag(\bm{p})\bigg]^{-1}\nonumber\\
  &\bigg[\Delta(\bm{p})\bigg(J\langle T_\tau J_{ab}\sigma_{\beta a}c^\dag_{\bm{p},\beta}\tilde{\vec{S}}c^\dag_{-\bm{p},b}\rangle+2\sum_{\bm{q},\bm{p}'}J_{ab}\tilde{U}c^\dag_{\sigma,\bm{q}}c_{\sigma,\bm{q}-\bm{p}'}c_{a,\bm{p}+\bm{p}'}c^\dag_{b,-\bm{p}}\nonumber\\
  &-\frac{J}{2}\sum_{\bm{q}}J_{ab}c_{a,\bm{p}}c^\dag_{\gamma,\bm{q}}c_{\gamma,\bm{q}}c^\dag_{b,-\bm{p}}\bigg)+(\mi p_n+\xi_p)\bigg(1-J\langle T_\tau\sigma_{\alpha\beta}c_{\bm{p},\beta}\tilde{\vec{S}}c^\dag_{\bm{p},\alpha}\rangle\nonumber\\
  &-2\sum_{\bm{q},\bm{p}'}\tilde{U}c^\dag_{\sigma,\bm{q}}c_{\sigma,\bm{q}+\bm{p}'}c_{a,\bm{p}-\bm{p}'}c^\dag_{a,\bm{p}}+\frac{J}{2}\sum_{\bm{q}}c_{a,\bm{p}}c^\dag_{\gamma,\bm{q}}c_{\gamma,\bm{q}}c^\dag_{a,\bm{p}}\bigg)\bigg]\,,
  \label{eq:solG}\\
  \mathcal{F}^\dag(\bm{p},\mi p_n)&=\bigg[p^2_n+\xi^2_p+\Delta(\bm{p})J_{ab}\epsilon_{ab}\Delta^\dag(\bm{p})\bigg]^{-1}\nonumber\\
  &\bigg[(\mi p_n-\xi_p)\bigg(J\langle T_\tau J_{ab}\sigma_{\beta a}c^\dag_{\bm{p},\beta}\tilde{\vec{S}}c^\dag_{-\bm{p},b}\rangle+2\sum_{\bm{q},\bm{p}'}J_{ab}\tilde{U}c^\dag_{\sigma,\bm{q}}c_{\sigma,\bm{q}-\bm{p}'}c_{a,\bm{p}+\bm{p}'}c^\dag_{b,-\bm{p}}\nonumber\\
  &-\frac{J}{2}\sum_{\bm{q}}J_{ab}c_{a,\bm{p}}c^\dag_{\gamma,\bm{q}}c_{\gamma,\bm{q}}c^\dag_{b,-\bm{p}}
\bigg)+J_{ab}\epsilon_{ab}\Delta^\dag(\bm{p})\big(1-J\langle T_\tau\sigma_{\alpha\beta}c_{\bm{p},\beta}\tilde{\vec{S}}c^\dag_{\bm{p},\alpha}\rangle\nonumber\\
  &-2\sum_{\bm{q},\bm{p}'}\tilde{U}c^\dag_{\sigma,\bm{q}}c_{\sigma,\bm{q}+\bm{p}'}c_{a,\bm{p}-\bm{p}'}c^\dag_{a,\bm{p}}+\frac{J}{2}\sum_{\bm{q}}c_{a,\bm{p}}c^\dag_{\gamma,\bm{q}}c_{\gamma,\bm{q}}c^\dag_{a,\bm{p}}\big)\bigg]\,.\label{eq:solF}
\end{align}
By inserting (\ref{eq:solF}) into
\begin{equation}   \Delta(\bm{p})=\Delta^\dag(\bm{p})=-4\sum_{\bm{q}}K\mathcal{F}^\dag(\bm{p}-\bm{q},\tau=0)=-4\sum_{\bm{q},p_n,q_n}K\mathcal{F}^\dag(\bm{p}-\bm{q},\mi p_n-\mi q_n)\,,
\end{equation}
we obtain the equation for the gap function, which is
\begin{align}
   &\Delta(\bm{p})=-4\sum_{\bm{p},p_n,q_n}K\bigg[(p_n-q_n)^2+\xi^2_p+J_{ab}\epsilon_{ab}\big(\Delta(\bm{p}-\bm{q})\big)^2\bigg]^{-1}\nonumber\\
   &\bigg[(\mi p_n-\mi q_n-\xi_{p-q})\bigg(J\langle T_\tau J_{ab}\sigma_{\beta a}c^\dag_{\bm{p}-\bm{q},\beta}\tilde{\vec{S}}c^\dag_{-\bm{p}+\bm{q},b}\rangle+2\sum_{\bm{q},\bm{p}'}J_{ab}\tilde{U}c^\dag_{\sigma,\bm{q}}c_{\sigma,\bm{q}-\bm{p}'}c_{a,\vec{p}+\vec{p}'}c^\dag_{b,-\vec{p}}\nonumber\\
   &-\frac{J}{2}\sum_{\bm{q}}J_{ab}c_{a,\bm{p}}c^\dag_{\gamma,\bm{q}}c_{\gamma,\bm{q}}c^\dag_{b,-\bm{p}}
\bigg)+J_{ab}\epsilon_{ab}\Delta(\bm{p}-\bm{q})\bigg(1-J\langle T_\tau\sigma_{\alpha\beta}c_{\bm{p}-\bm{q},\beta}\tilde{\vec{S}}c^\dag_{\bm{p}-\bm{q},\alpha}\rangle\nonumber\\
   &-2\sum_{\bm{q},\bm{p}'}\tilde{U}c^\dag_{\sigma,\bm{q}}c_{\sigma,\bm{q}+\bm{p}'}c_{a,\bm{p}-\bm{p}'}c^\dag_{a\bm{p}}+\frac{J}{2}\sum_{\bm{q}}c_{a,\bm{p}}c^\dag_{\gamma,\bm{q}}c_{\gamma,\bm{q}}c^\dag_{a,\bm{p}}\bigg)
\big)\bigg]\,.
\end{align}
We define the excitation energy of the superconductor as
\begin{equation}
  E_{p-q}=\sqrt{\xi^2+J_{ab}\epsilon_{ab}(\Delta(\bm{p}-\bm{q}))^2}\,.
\end{equation}
 The summation over $\mi(p_n-q_n)$ is evaluated by the contour integral
\begin{equation}
   \oint \frac{\md z}{2\pi \mi}n_F(z)\frac{\Delta^\dag(\bm{p}-\bm{q})}{z^2-E^2_{p-q}}\,,\,\oint \frac{\md z}{2\pi \mi}n_F(z)\frac{z-\xi_{p-q}}{z^2-E^2_{p-q}}\,,
\end{equation}
and the poles of Fermi distribution $n_F(z)=\frac{1}{n^{\beta z}+1}$ give the summation over $z=\mi(p_n-q_n)$. Since $\frac{\Delta}{2E_p}\frac{\me^{-\beta E_p}-\me^{\beta E_p}}{2+\me^{-\beta E_p}+\me^{\beta E_p}}=\frac{\Delta}{2E_p}\tanh(\frac{\beta E_p}{2})$, now the gap function is
\begin{align}
  \Delta(\bm{p})=\sum_{\bm{q}}f(\bm{q})&=4\sum_{\bm{q}}\bigg[-KJ_{ab}\epsilon_{ab}\bigg(1-J\langle T_\tau\sigma_{\alpha\beta}c_{\bm{p}-\bm{q},\beta}\tilde{\vec{S}}c^\dag_{\bm{p}-\bm{q},\alpha}\rangle\nonumber\\
  &-2\sum_{\bm{q},\bm{p}'}\tilde{U}c^\dag_{\sigma,\bm{q}}c_{\sigma,\bm{q}+\bm{p}'}c_{a,\bm{p}-\bm{p}'}c^\dag_{a,\bm{p}}(0)+\frac{J}{2}\sum_{\bm{q}}c_{a,\bm{p}}c^\dag_{\gamma,\bm{q}}c_{\gamma,\bm{q}}c^\dag_{a,\bm{p}}\bigg)
\big)\tanh\big(\frac{\beta E_{p-q}}{2}\big)\frac{\Delta(\bm{p}-\bm{q})}{2E_{p-q}}\nonumber\\
                       &-K\bigg(J\langle T_\tau J_{ab}\sigma_{\beta a}c^{\dag}_{\bm{p}-\bm{q},\beta}\tilde{\vec{S}}c^\dag_{-\bm{p}+\bm{q},b}\rangle+2\sum_{\bm{q},\bm{p}'}J_{ab}\tilde{U}c^\dag_{\sigma,\bm{q}}c_{\sigma,\bm{q}-\bm{p}'}c_{a,\bm{p}+\bm{p}'}c^\dag_{b,-\bm{p}}\nonumber\\
                       &-\frac{J}{2}\sum_{\bm{q}}J_{ab}c_{a,\bm{p}}c^\dag_{\gamma,\bm{q}}c_{\gamma\bm{q}}c^\dag_{b,-\bm{p}}
\bigg)\bigg(E_{p-q}-\xi_{p-q}\tanh\big(\frac{\beta E_{p-q}}{2}\big)\bigg)\frac{1}{2E_{p-q}}\bigg]\,.
\end{align}
It is convenient to change the summation to an integration
\begin{equation}
  \sum_{\bm{q}}f(\bm{q})=\int\frac{\md^3q}{(2\pi)^3}f(\bm{q})=N_F\int^{\omega_D}_{-\omega_D}\md\xi f(\xi)\,,
\end{equation}
where we have approximately substituted the constant $N_f$ for density of states near the Fermi surface. Taking the zero temperature limit $\beta=1/T\rightarrow\infty$, we obtain
\begin{align}
   \Delta(\bm{p})&=-4KN_F\bigg[J_{ab}\epsilon_{ab}\bigg(1-J\langle T_\tau\sigma_{\alpha\beta}c_{\bm{p}-\bm{q},\beta}\tilde{\vec{S}}c^\dag_{\bm{p}-\bm{q},\alpha}\rangle-2\sum_{\bm{q},\bm{p}'}\tilde{U}c^\dag_{\sigma,\bm{q}}c_{\sigma,\bm{q}+\bm{p}'}c_{a,\bm{p}-\bm{p}'}c^\dag_{a,\bm{p}}\nonumber\\
                  &+\frac{J}{2}\sum_{\bm{q}}c_{a,\bm{p}}c^\dag_{\gamma,\bm{q}}c_{\gamma,\bm{q}}c^\dag_{a,\bm{p}}\bigg)\frac{\Delta}{2}\ln(\xi+\sqrt{J_{ab}\epsilon_{ab}\Delta^2+\xi^2})|^{\omega_D}_{-\omega_D}+\bigg(J\langle T_\tau J_{ab}\sigma_{\beta a}c^{\dag}_{\bm{p}-\bm{q},\beta}\tilde{\vec{S}}c^\dag_{-\bm{p}+\bm{q},b}\rangle\nonumber\\
                  &+2\sum_{\bm{q},\bm{p}'}J_{ab}\tilde{U}c^\dag_{\sigma,\bm{q}}c_{\sigma,\bm{q}-\bm{p}'}c_{a,\bm{p}+\bm{p}'}c^\dag_{b,-\bm{p}}-\frac{J}{2}\sum_{\bm{q}}J_{ab}c_{a,\bm{p}}c^\dag_{\gamma,\bm{q}}c_{\gamma,\bm{q}}c^\dag_{b,-\bm{p}}\bigg)\omega_D\bigg]\,.\label{eq:gap2}
\end{align}
Since $\Delta$ is constant and $\ln(\xi+\sqrt{J_{ab}\epsilon_{ab}\Delta^2+\xi^2})|^{\omega_D}_{-\omega_D}\approx2\ln(\frac{2\omega_D}{\sqrt{J_{ab}\epsilon_{ab}}\Delta})$, (\ref{eq:gap2}) leaves the equation for the energy gap
\begin{align}
  \Delta&=-4KN_F(A\Delta\ln(\frac{2\omega_D}{\sqrt{J_{ab}\epsilon_{ab}}\Delta})+\omega_D B)\,,\label{eq:gap}\\
  A&=J_{ab}\epsilon_{ab}\bigg(1-J\langle T_\tau\sigma_{\alpha\beta}c_{\bm{p}-\bm{q},\beta}\tilde{\vec{S}}c^\dag_{\bm{p}-\bm{q},\alpha}\rangle\rangle-2\sum_{\bm{q},\bm{p}'}\tilde{U}c^\dag_{\sigma,\bm{q}}c_{\sigma,\bm{q}+\bm{p}'}c_{a,\bm{p}-\bm{p}'}c^\dag_{a,\bm{p}}+\frac{J}{2}\sum_{\bm{q}}c_{a,\bm{p}}c^\dag_{\gamma,\bm{q}}c_{\gamma,\bm{q}}c^\dag_{a,\bm{p}}\bigg)\,,\label{eq:A}\\
  B&=J\langle T_\tau J_{ab}\sigma_{\beta a}c^{\dag}_{\bm{p}-\bm{q},\beta}\tilde{\vec{S}}c^\dag_{-\bm{p}+\bm{q},b}\rangle+2\sum_{\bm{q},\bm{p}'}J_{ab}\tilde{U}c^\dag_{\sigma,\bm{q}}c_{\sigma,\bm{q}-\bm{p}'}c_{a,\bm{p}+\bm{p}'}c^\dag_{b,-\bm{p}}-\frac{J}{2}\sum_{\bm{q}}J_{ab}c_{a,\bm{p}}c^\dag_{\gamma,\bm{q}}c_{\gamma,\bm{q}}c^\dag_{b,-\bm{p}}\,.\label{eq:B}
\end{align}
We set $\omega_D=1$ to fit the gap, and choose a small correction for spin $S=\langle T_\tau\sigma c^\dag Sc^\dag\rangle$. In BCS theory (i.e. $A=1,B=0$), the energy gap for $K=-\frac{1}{4}V_0$ is $\Delta=2\omega_D \rm{e}^{-1/V_0 N_F}$ at zero temperature $(V_0>0)$. $-V_0$ is the attractive and constant potential in BCS theory. Equation (\ref{eq:gap}) could be numerically calculated and the result is shown in Figure \ref{fg:Figure 1}. Here we have neglected the effect of B term due to the following analysis on $T_c$. We could conclude that the energy gap in the ``tetrahedron" model is higher than the BCS energy gap represented by red line when $U=K=-J/2$.
Furthermore, we know $\Delta(T=T_c)=0$ at the transition temperature $T_c$. Then, (\ref{eq:gap2}) becomes
\begin{equation}
  1=-4KN_F\int^{\omega_D}_{-\omega_D}\md\xi\bigg[\frac{A}{2\xi}\tanh\big(\frac{\xi}{2T}\big)+\frac{B}{2\Delta}\bigg(1-\tanh\big(\frac{\xi}{2T}\big)\bigg)\bigg]\,.\label{eq:gap3}
\end{equation}
Using the Euler integral formula, we obtain the transition temperature as follow
\begin{equation}
  T_c=1.13\omega_D \me^{1/(4KN_FA)}\,.
\end{equation}
 Since we have required that (\ref{eq:gap3}) must be regular, it yields
\begin{equation}
  A=J_{ab}\epsilon_{ab}\big(1-J\langle T_\tau\sigma_{\alpha\beta}c_{\bm{p}-\bm{q},\beta}\tilde{\vec{S}}c^\dag_{\bm{p}-\bm{q},\alpha}\rangle\big)\,,\,B=0\,.
\end{equation}
  We notice that the critical temperature is $T_c=1.13\omega_D \rm{e}^{-1/V_0N_F}$ in the BCS theory. While our solution of $T_c$ is modified by $K$ and $S$. We plot the the transition temperature $T_c$ as a function of the coupling $K$ of the SYK-like term in Figure \ref{fg:Figure 2}. The transition temperature decrease as $K$ increase. $K$ is the SYK-like coupling. As to the energy gap, the transition temperature diverges as $K$ goes from negative to zero.
Now we have both the energy gap and the transition temperature. The ratio of these two results is $\frac{2\Delta}{T_c}=3.5$ in the BCS theory. When $\frac{2\Delta}{T_c}>3.5$, it is the case of strong coupling. As we know, the energy gap and the critical temperature are dependent on the coupling $V_0$, while $\frac{2\Delta}{T_c}$ is independent on $V_0$ in the BCS theory. Since the ratio $\frac{2\Delta}{T_c}$ is dependent on the coupling $K$ in the  ``tetrahedron'' model, it is interesting to show the numerical evaluation of $\frac{2\Delta}{T_c}$ in Figure \ref{fig:Figure 3}.
According to the numerical evaluation, we conclude that the gap ratio could be higher than the one in BCS theory in the case of $U=K=-J/2$.  When $S=0.05,K=-J/2=-5$, which is higher than the gap ratio in BCS theory, we have $\frac{2\Delta}{T_c}\thickapprox 5$. Then, the gap ratio decreases as $K$ increases but $S$ decreases in the case of $U=K=-J/2$. However, in the case of $U=K=J/2$, the gap ratio could not exceed the one in BCS theory. If $S$ vanishes, the ratio $\frac{2\Delta}{T_c}=3.5$ in the ``tetrahedron'' model ($K<0$) is exactly the same as the ratio in the BCS theory ($K=1$). In other words, the ratio is independent of the coupling $K$ in such case.

\section{Conclusion and discussion}

In this paper, we attempt to understand the pairing mechanism of high-temperature superconductivity, which extends the BCS theory to strong coupling. For this purpose, SP(M) singlet pairing operator is proposed in an SYK-like model. Then equations for the correlation functions are derived. Our analysis shows how the superconducting gap, the transition temperature and the their ratio change with the coupling $K$ and spin $\langle T_\tau\sigma c^\dag Sc^\dag\rangle$. When $U=K=-J/2$, the ratio $\frac{2\Delta}{T_c}>3.5$. This result indicates that the SYK-like model relates to the case of strong coupling. The behavior of this model at strong coupling limit beyonds the scope of this paper. We also leave the ratio of susceptibility and specific heat to a future study. Specially, the energy gap, the transition temperature and the ratio $\frac{2\Delta}{T_c}$ could return to the BCS theory if $<\langle T_\tau\sigma c^\dag Sc^\dag\rangle=0$.

The interaction term of our model is not random, but it demonstrates features of strange metal. There is other system with non-random interaction. It becomes non-Fermi liquid metal with a superconducting instability \cite{Phillips:2019qva}. Actually, the single particle Green's function with large component tensor is identical to the disordered averaged Green's function of the SYK models \cite{WCJYX,Sachdev:2015efa}. The full Green's function and the current vertex of the translational invariant model with random interaction terms could be solvable in the large $N$ limit \cite{Chowdhury:2018sho}. Thus, in the SYK model at large $N$ limit, the quantum contribution to (\ref{eq:A})(\ref{eq:B}) of the rank-3 tensor model can be summed analytically.

 Our calculation may be not applied in the large $N$ limit, due to the long range interaction between lattices. Although we could not generalize our calculations to large $N$ limit, enhancement of the gap ratio is still seen in the model at large $N$ limit \cite{Patel:2018rjp}. Two lattice models are proposed with on-site SYK interactions exhibiting a transition from an IM to an s-wave superconductor in \cite{Patel:2018rjp}. In some holographic superconductors, the gap ratio increases as well \cite{Hartnoll:2008vx}. On the other hand, in \cite{WCJYX} it is also argued that the correction to the NFL solution in this model is suppressed rapidly with increasing $N$. Therefore, our results without so large $N$ show a qualitative agreement.

\section*{Acknowledgements}
  We would like to thank Shao-Kai Jian and Shi-Ping Zhou for valuable discussions. The study was partially supported by NSFC China (Grant No. 11805117 and Grant No. 11875184).

\end{document}